# Tunable topological phononic crystals

Ze-Guo Chen and Ying Wu

Division of Computer, Electrical and Mathematical Sciences and Engineering, King Abdullah University of Science and Technology, Thuwal 23955-6900, Saudi Arabia.

**Abstract**

Topological insulators, first observed in electronic systems, have inspired many analogues in photonic and phononic crystals in which remarkable one-way propagation edge states are supported by topologically nontrivial bandgaps. Such bandgaps can be achieved by breaking the time-reversal symmetry to lift the degeneracy associated with Dirac cones at the corners of the Brillouin zone. Here, we report on our construction of a phononic crystal exhibiting a Dirac-like cone in the Brillouin zone center. We demonstrate that simultaneously breaking the time-reversal symmetry and altering the geometric size of the unit cell result in a topological transition that is verified by the Chern number calculation and edge mode analysis. The topology of the bandgap is tunable by varying both the velocity field and the geometric size; such tunability may dramatically enrich the design and use of acoustic topological insulators.

Topological order, a mathematic concept of conserved properties under continuous deformations, was first introduced as a new property of the full energy band along with the discovery of the integer quantum Hall effect (QHE) [1,2]. The topological invariant of a two-dimensional (2D) energy band is called the Chern number, a quantity that characterizes the quantized collective behavior of the wave functions of the band. When the sum of the Chern numbers of all the bands below a bandgap is not zero, a so-called "topological insulator" emerges. A fascinating property of these topological insulators is that they support a one-way propagation edge state without back scattering on their surfaces. Though topological insulators were first observed in electronic systems [3], the concept has been successfully extended to other wave systems because of similarities in the band structures of those systems. The photonic [4-8], elastic [9-17] and mechanical [18-20] analogues of electronic topological insulators have been proposed.

One possible way to achieve a nonzero Chern number would be to capitalize on the physics of the QHE and break the time-reversal ($\mathcal{T}$) symmetry, which is feasible in photonic crystals by introducing magnetic fields. However, breaking the $\mathcal{T}$ symmetry is fundamentally difficult in acoustic systems because ordinary acoustic materials generally conserve their $\mathcal{T}$ symmetry. Very recently, Alu et al. tackled this problem by introducing a rotation flow in the background field to break the $\mathcal{T}$ symmetry in an acoustic system [21]. Later, their solution was generalized to phononic crystals to achieve acoustic topological insulators [9,10,14].

Most of the progress in this area [9,10,14] was made with lattices with $C_{6v}$ symmetry in which there exists a Dirac cone, protected by the $\mathcal{T}$ symmetry, at the $K$ point of the Brillouin zone. The existence of the Dirac cone is attributed to the symmetry of the lattice; it is not affected by the geometric size of the inclusions in the crystal [22]. Because of the linear dispersion of the Dirac cone, breaking the $\mathcal{T}$ symmetry may open a nontrivial bandgap and introduce edge states[23]. Recently, another kind of linear dispersion, namely a Dirac-like cone, was found at the center of the Brillouin zone of lattices with both $C_{6v}$ and $C_{4v}$ symmetries [24,25]. Different from Dirac cones, Dirac-like cones arise from accidental degeneracy and are sensitive to the geometric size of the inclusion. They are therefore protected by both the $\mathcal{T}$ symmetry and the particular geometric parameters of the crystal. A comprehensive understanding of the degrees of freedom needed to modulate the topological invariant would provide a physical picture of the mechanism of the bandgap opening at the Brillouin zone center, further expanding our ability to tailor acoustic waves. We therefore wondered what the consequences would then if breaking the $\mathcal{T}$ symmetry and altering the geometric size of the crystal occurred simultaneously.

To answer our question, we constructed a 2D phononic crystal with $C_{4v}$ symmetry. By carefully adjusting the size of the crystal's building block, we found a Dirac-like cone in its band structure associated with a triply degenerate state at the Brillouin zone center. We observed that the degeneracy can be lifted by altering the geometrical parameters of the inclusion or by introducing airflow, and both ways can introduce a bandgap at the frequency of the degeneracy. We developed a tight-binding model to systematically study the underlying mechanism of the bandgap opening, which revealed that altering the geometry of the inclusion is equivalent to changing the coupling coefficients between the neighboring lattices and that introducing airflow indicates that the $\mathcal{T}$ symmetry has been broken. By calculating the Chern numbers of the bands, we found that the induced bandgaps exhibit different topologies. Changing the geometric size of the inclusion does not give rise to nontrivial bandgaps, whereas introducing airflow does. We further explored the system by combining together the two ways of lifting the degeneracy. A somewhat unexpected result is that breaking the $\mathcal{T}$ symmetry is not sufficient to yield a nontrivial bandgap if the geometric size of the inclusion is below a certain critical point. This critical point is called the topological transition point and its value depends on the intensity of the introduced airflow. Based on this interesting finding, we demonstrate, with numerical experiments, the existence of a tunable state in the phononic crystal. When the geometric size of the inclusion or the intensity of the airflow exceed the topological transition point, the state changes from a localized state (attributed to the trivial bandgap) into a topology-protected one-way edge state (a result of the nontrivial bandgap) that is immune to back scattering.

The 2D phononic crystal considered here is composed of a square array of acoustic waveguides. The unit cell is a hollow ring attached by four subwavelength rectangular waveguides as illustrated in Fig. 1(a). The lattice constant is $a = 2m$ and the inner and outer radii of the ring are $r_0 = 0.35m$ and $r_1 = 0.5m$, respectively. The width of the rectangular waveguide is $d$, which can be adjusted. Inside the ring, the air flows counterclockwise (we assume that the flow is confined within the ring). The acoustic wave propagation in such a structure is protected by the $\mathcal{T}$ symmetry if there is no airflow, but it would behave according to the following equation if an irrotational airflow were introduced [26]:

$$-\frac{\rho}{c^2}i\omega(i\omega\phi+\vec{v}\cdot\nabla\phi)+\nabla\cdot(\rho\nabla\phi-\frac{\rho}{c^2}(i\omega\phi+\vec{v}\cdot\nabla\phi)\vec{v})=0, \quad (1)$$

where $\phi$ represents the velocity potential, $\rho$ and $c$ denote the mass density and the velocity of sound in air, $\omega$ is the angular frequency of the acoustic wave, and $\vec{v}$ is the velocity field of the airflow.

We start with a phononic crystal without the introduction of airflow. The band structure calculation performed by COMSOL Multiphysics (a finite-element software) shows that there are three eigenmodes that degenerate at the $\Gamma$ point when the width of the waveguide ($d$) equals $0.07336m$. Their corresponding pressure field distributions are shown in Fig. 1(b); two of these distributions share a similar pattern and are marked in the figure as $\varphi_{px}$ and $\varphi_{py}$ and the third one is marked as $\varphi_s$. We note that these eigenmodes are calculated with the periodic boundary condition imposed. If a unit cell is placed in the free space, modes with similar symmetries, denoted as $\Phi_{px}$, $\Phi_{py}$ and $\Phi_s$, also exist. By taking the symmetry of these free-space modes into account, we develop a tight-binding model to describe the dispersion relation associated with these modes [27,28]. The kernel of the Hamiltonian is expressed as follows,

$$H = \begin{bmatrix} E_s + 2t_x^{11}(\cos k_x a + \cos k_y a) & 2it_x^{12}\sin k_x a & 2it_x^{12}\sin k_y a \\ -2it_x^{12}\sin k_x a & E_{px} + 2t_x^{22}\cos k_x a + 2t_y^{22}\cos k_y a & 0 \\ -2it_x^{12}\sin k_y a & 0 & E_{py} + 2t_y^{22}\cos k_x a + 2t_x^{22}\cos k_y a \end{bmatrix} \quad (2)$$

where $E_s$, $E_{px}$ and $E_{py}$ are the on-site energy of the rings and $t_m^{ij} = \langle \Phi_i(\vec{r}) | H | \Phi_j(\vec{r} + \vec{r}_m) \rangle$ represents different types of first-neighbor coupling coefficients between the above-mentioned free-space modes, where $\Phi_i (i=1,2,3)$ correspond to $\Phi_s$, $\Phi_{px}$ and $\Phi_{py}$, respectively. $\vec{r}_m (m=x, y)$ are lattice vectors. Here, Eq. (2) has been simplified by considering the symmetry, for example $t_x^{22} = t_y^{33}$. The coupling coefficient is mainly contributed by the wave functions inside the waveguide, and therefore is a function of the width of the waveguide. At $\Gamma$ point ($k_x = k_y = 0$), Eq. (2) becomes a diagonal matrix, and its diagonal terms are nothing but the eigenvalues of the Hamiltonian, which are $E_s + 4t_x^{11}$, $E_{px} + 2t_x^{22} + 2t_y^{22}$ and $E_{py} + 2t_x^{22} + 2t_y^{22}$. These eigenvalues are proportional to the eigenfrequencies of the $\varphi_s$, $\varphi_{px}$ and $\varphi_{py}$ modes, respectively. The on-site energy does not change because the size of the ring is fixed. The coupling coefficients are proportional to the width of the rectangular waveguide, because the waveguide is narrow enough and it only supports the fundamental mode. We can therefore expect that the eigenvalues of the Hamiltonian (or the eigenfrequencies of the three modes) should depend linearly on the width of the waveguide. Such linear behavior is verified by numerical simulations. We use COMSOL to calculate the band structure of the same phononic crystal but with different rectangular waveguide sizes and plot in Fig. 1(c) the eigenfrequencies of $\varphi_s$, $\varphi_{px}$ and $\varphi_{py}$. (The detailed band structures for different $d$ are illustrated in Fig. 2(c)-2(d).) Due to the symmetry of the lattice, $\varphi_{px}$ and $\varphi_{py}$ always degenerate and therefore they share the same eigenfrequency, which is indicated by the red line in Fig. 1(c). The black line indicates the eigenfrequency of the $\varphi_s$ mode. The black and red lines intersect at $d = d_0 = 0.07336m$ where the accidental degeneracy of the three states occurs and the eigenvalues of the Hamiltonian become identical. In fact, $d_0$ also indicates a transition point because both the tight-binding model and the simulation of the real systems show the existence of a band gap only when $d < d_0$. This band gap is topologically trivial because the $\mathcal{T}$ symmetry is preserved.

When airflow is introduced, the $\mathcal{T}$ symmetry breaks. Here, for simplicity but without loss of generality, we choose the following velocity field distribution [10,14]:

$$\vec{v}(x, y) = (\frac{-vy}{\sqrt{x^2 + y^2}}, \frac{vx}{\sqrt{x^2 + y^2}}) = v\vec{e}_\theta, \quad (3)$$

where $\vec{e}_\theta$ is the azimuthal unit vector along the counterclockwise direction and $v$ is the amplitude of the velocity field. The advantage of choosing such a velocity field distribution is because the radial component of $\vec{v}$ is a constant. The velocity field is invariant under time reversal, and it is not difficult to see that the system, as well as its describing equation (Eq. (1)), is no longer symmetric to time reversal. As a result, the degeneracy of the on-site energy, $E_{px}$ and $E_{py}$, is lifted as is the degeneracy of $\varphi_{px}$ and $\varphi_{py}$. From the field patterns shown in Fig. 1(b), where there is no airflow, we find that the $\varphi_{px}$ and $\varphi_{py}$ modes are mainly from the resonance inside the ring when the circumference of the ring approximately equals the wavelength, whereas the $\varphi_s$ mode is mainly from the resonance inside the rectangular waveguide. Intuitively, circulating airflow would not affect the eigenfrequency of $\varphi_s$ by much, but it would affect the eigenfrequencies of $\varphi_{px}$ and $\varphi_{py}$. According to the superposition principle, the acoustic wave circulates inside the ring at different velocities, $c+v$ and $c-v$, which leads to the splitting of the resonance frequencies as follows: $\omega_\pm = (c \pm v)/R_{av}$, where $R_{av} = (r_0 + r_1)/2$ is the average of the inner and outer radii and $\omega_0 = c/R_{av}$ is the eigenfrequency of the degenerated modes, $\varphi_{px}$ and $\varphi_{py}$, without the airflow [21]. It is obvious that the splitting of the resonance frequencies of $\varphi_{px}$ and $\varphi_{py}$ is linear in the external velocity field, $v$. In Fig. 1(d), we plot the COMSOL-calculated eigenfrequencies of the three modes as functions of the velocity field. This plot indeed shows the linear behavior in the splitting of the eigenfrequencies of $\varphi_{px}$ and $\varphi_{py}$ and the almost constant eigenfrequency of $\varphi_s$ as $v$ increases, supporting our analysis.

Altering the size of the rectangular waveguide and introducing the airflow would affect the eigenfrequencies of the three modes and change the properties of the degeneracy in different ways. Of particular interest are the effects on the properties of the band structure played by these two processes. In Figs. 2(b)-2(d), we plot the COMSOL-calculated band structures for phononic crystals with $d_0 = 0.07336m$, $d_1 = 0.04m$, and $d_2 = 0.1m$, respectively. The blue curves are obtained without an additional velocity field, i.e., $v = 0$. As mentioned above, accidental degeneracy between $\varphi_s$ and $\varphi_{px}$ ($\varphi_{py}$) is achieved at $d = d_0$ and linear dispersion is observed in the center of the Brillouin zone. This linear dispersion is also consistent with the tight-binding model proposed earlier. Meanwhile two double-degenerated points are found in the corner of the Brillouin zone, i.e., the $M$ point. If $d$ deviates from $d_0$, the accidental degeneracy at the $\Gamma$ point no longer exists and the branch associated with $\varphi_s$ does not touch the branches associated with $\varphi_{px}$ and $\varphi_{py}$. When $d = d_1 < d_0$, a band gap opens in the $138.2Hz$ to $141.3Hz$ frequency range, while no band gap appears when $d = d_2 > d_0$ even though the accidental degeneracy is lifted. We also observe that the doubly degenerated modes at the $M$ point degenerate as $d$ changes, because the degeneracy arises from the symmetry of the lattice and is thus deterministic.

We introduce an airflow at $v = 10m/s$ into the systems and plot the corresponding band structures as red curves in Figs. 2(b)-2(d). As expected, the branch associated with $\varphi_s$ is mostly unaffected, and the branches associated with $\varphi_{px}$ and $\varphi_{py}$ do not degenerate at the $\Gamma$ point. Consequently, complete band gaps open for the cases when $d = d_0$ and $d = d_2$. For the case of $d = d_1$, introducing the airflow only changes the behavior of the band structure, but it does not open a complete band gap. Meanwhile, the degeneracies at the $M$ point are also lifted for all the three cases because of the broken $\mathcal{T}$ symmetry. The changes in the band structures of all three cases after introducing the airflow suggest that there would be a topological transition point at $d = d_t$ for a particular $v$. The lifted degeneracy would (or would not) lead to a new bandgap when $d \geq d_t$ (or $d < d_t$). It is not difficult to see in Fig. 2(a) that at $d_t$, the eigenfrequency of $\varphi_s$ is the same as that of $\varphi_{py}$.

The bandgap can be characterized by a topological invariant called a "gap Chern number". A general way to calculate this invariant is to sum the Chern numbers, which are expressed by the following equation, of all bands below the bandgap:

$$C = \frac{i}{2\pi} \sum_n \int_{BZ} \nabla_{\vec{k}} \times (\langle u_n(\vec{k}) | \nabla_{\vec{k}} | u_n(\vec{k}) \rangle) d^2\vec{k} \quad , \tag{4}$$

where $u_n(\vec{k})$ is the Bloch function of the $n$th band at $\vec{k}$. It is possible to substitute the numerically simulated eigenstates on the band into Eq. (4) to compute the Chern number, but this is computationally challenging. It is also possible to write out the Hamiltonian including the airflow and solve for the eigenvectors and plug them into Eq. (4). Here, the introduced flow field, $\vec{v}$, can be viewed as the magnetic vector potential field, $\vec{A}$, employed in the Haldane model [9,10,29]. However, there is a fundamental difference because the magnetic vector potential field has gauge freedom whereas the flow field does not. Furthermore, the vector potential is dependent on the frequency [9], which makes the calculation of the Chern number over the whole Brillouin zone difficult because the Hamiltonian is valid only near the high symmetry points. Fortunately, only the region near the broken degeneracy points generates a nonvanishing "Berry flux," which contributes to the band's Chern number [30], indicating that we can construct Hamiltonians at the points of degeneration to determine the Chern numbers of the four bands. This method has been utilized in determining the Chern number of the bands connected by Dirac cones in phononic crystals with broken $\mathcal{T}$ symmetry and with $C_{3v}$ or $C_{6v}$ symmetry [10]. To the best of our knowledge, there is no explicit result on the Chern number for the case when a three-folded accidental degeneracy is lifted by breaking $\mathcal{T}$ symmetry. Recently, the Chern number was found to be constrained by the symmetry properties of the eigenstates at high symmetry points [31].

The point group of $\vec{k}$ at the $\Gamma$ point becomes $C_4$ when the airflow is introduced. We consider all the $C_m$-invariant points below the bandgap in a $C_n$-invariant insulator for each $m$ dividing $n$. The gap Chern number modulo, $n$, is related to the eigenvalues of corresponding $C_m$ operators of all these points:

$$i^C = \prod_j \xi_j(\Gamma)\xi_j(M)\zeta_j(Y) \quad , \tag{5}$$

where $j$ is the label of bands under the bandgap and $\xi$ and $\zeta$ are the respective eigenvalues of the $\hat{C}_4$ and $\hat{C}_2$ operators on the eigenstates at high symmetry points. Table (1) presents the calculated results from Eq. (5), indicating the various topological properties of the bandgaps. The bandgaps with different Chern numbers are highlighted in different colors in Fig. 2(a) and marked by numbers in Figs. 2(b)-2(d). From Fig. 2(a), we found that the Chern number remains zero when $d$ is smaller than $d_t$ even with broken $\mathcal{T}$ symmetry and it changes to one when $d$ is larger than $d_t$. When $d$ becomes even larger, a band inversion across the point $d = d_i$ also exists, but it does not contribute to the gap Chern number. This clearly suggests the existence of a topological transition that occurs by modulating the geometric parameters of the inclusion in a system with broken $\mathcal{T}$ symmetry, which renders additional freedom to manipulate the gap Chern number in the 2D case.

Insulators with nonzero gap Chern numbers are topologically nontrivial. A novel property in such insulators is the presence of gapless edge state between gaps with different topological invariants [1]. The band structure of an $8\times 1$ supercell with $d = 0.1m$ and $v = 10m/s$ is calculated to confirm the existence of such gapless edge states. This supercell is infinite along the $x$-direction and is terminated by rigid boundaries (topologically trivial) in the $y$-direction. Figure 3(a) shows the band structure of such a supercell. It exhibits one one-way edge mode, which agrees with the gap Chern numbers. The dispersion relation highlighted in blue (red), marked as A (B), is associated only with modes having a negative (positive) group velocity. At a given frequency, the modes corresponding to the A and B branches are confined on the opposite edges of the supercell as shown in Fig. 3(b). The A and B branches support modes bounded on the top and bottom edges, respectively. This property leads to one-way propagation at the edge. At some particular frequencies, located inside the gap region of the bulk mode, there exists only an edge mode, implying that there is no backward scattering into the bulk. The features exhibited in Figs. 3(a) and 3(b) ensure the existence of topologically protected one-way propagation of the edge modes.

To demonstrate the existence of acoustic one-way edge mode propagation, we perform finite element simulations of some finite-sized samples. The first sample is composed of $8\times 24$ unit cells as shown in Fig. 4(a). The upper, bottom and right boundaries of the sample are hard walls that can be treated as insulators with zero Chern numbers. A plane wave radiation condition is set on the left boundary. A point source with frequency $139 Hz$, a common bandgap frequency for various widths, $d$, is located in the bottom boundary. Also in the bottom boundary is a defect that is introduced by removing the airflow inside the ring, which is marked as a green circle in Fig. 4(a). The pressure field distribution unambiguously shows that the sound wave propagates counterclockwise and circumvents the defect without backscattering.

Changing the topological invariant by varying the geometry also allows us to create the interface states. Another finite-sized sample with $8\times 25$ lattices is presented in Fig. 4(b). Its upper half contains $4\times 25$ lattices in which the width of the waveguide is $d_1$; its lower half contains the other lattices in which the width of the waveguide is changed into $d_2$. A point source is located on the interface between the upper and lower halves. The frequency remains $139 Hz$. According to our previous analysis, this frequency is located inside the bandgaps of both the upper and lower lattices. The bandgap is trivial for the upper lattices and nontrivial for the lower lattices. This means that no energy would penetrate into the upper lattices and that there should be an interface state. The pressure field shown in Fig. 4(b) indeed demonstrates the existence of the interface state and the unidirectional behavior of the propagation of the sound wave due to the different topologies of the bandgaps of the upper and lower lattices.

As supported by the results shown in Fig. 4, the topology of the bandgap can be tuned by changing the geometry of the phononic crystal. On the other hand, according to the tight-binding model and symmetry analysis presented earlier, varying the velocity field is another effective way to tune the topology of the bandgap. In Fig. 5(a), we plot the eigenfrequencies of $\varphi_s$ and $\varphi_{py}$ as functions of the velocity field for a lattice with a fixed waveguide width, i.e., $d = 0.065m$. Similar to Fig. 3(a), we use purple to indicate the trivial bandgaps and blue to indicate the nontrivial bandgaps. We compare the sound wave propagation in two finite-sized samples. They share the same geometric configuration and are excited by the same source, but they are exposed to distinct airflows. The one with velocity field $v = 5m/s$ (Fig. 5(b)) exhibits typical trivial bandgap behavior as the acoustic pressure field is almost localized around the source, whereas the other one with velocity field $v = 15m/s$ (Fig. 5(c)) exhibits one-way edge states propagating counterclockwise.

In conclusion, we report on our design of a tunable phononic crystal that exhibits topologically non-trivial bandgaps by breaking the $\mathcal{T}$ symmetry and modulating the geometric parameters of the inclusion. The mechanisms that lead to bandgap opening by breaking the $\mathcal{T}$ symmetry and by varying the geometric parameters of the inclusion as well as their interplay are studied systematically by using a tight-binding model, a rigorous symmetry analysis and numerical simulations. We find a topological transition point that is related to both $\mathcal{T}$ symmetry and the geometric size of the inclusion, which suggests that the topology can be changed by tuning the strength of the velocity field and/or the size. The transition from a localized state to a robust one-way propagated edge mode is verified by our numerical experiments. Our findings could inspire new designs of acoustic topological materials, which should improve applications that require one-way propagation.

Acknowledgements:

The authors would like to thank X. Ni, X. C. Sun, X. J. Zhang for stimulating discussions. The work described here is supported by King Abdullah University of Science and Technology.

Figure captions:

FIG.1 (a) Schematics of a unit cell of our phononic crystal. The ring is connected by rectangular waveguides. (b) Pressure field distributions of three Bloch eigenstates at $\Gamma$ point, marked as $\varphi_s$, $\varphi_{px}$ and $\varphi_{py}$. Red and blue indicate the positive and negative maxima. The lower inset illustrates the reciprocal lattice. (c) The eigenfrequency of the eigenstates varies as a function of $d$ when no airflow is introduced. The black and red curves correspond to the $\varphi_s$ and $\varphi_{px}$ ($\varphi_{py}$) eigenstates, respectively. The bandgap region is colored purple, indicating that $d_0$ is a transition point. There is a bandgap when $d < d_0$ and there is no bandgap when $d > d_0$. (d) The eigenfrequency of $\varphi_s$, $\varphi_{px}$ and $\varphi_{py}$ splits as a function of the velocity field of the induced airflow.

FIG. 2. Changes in the band structures of our phononic crystals with constant airflow velocity, $v = 10 m/s$ and various $d$. (a) The eigenfrequency of the eigenstates varies as a function of $d$, black, blue and red curves correspond to the $\varphi_s$, $\varphi_{px}$ and $\varphi_{py}$ modes, respectively. The purple and blue areas indicate the region of the bandgap with different topological invariants, where $d_t$ is the topological transition point. $d_i$ corresponds to a band inversion point, but there is no topological transition. (b) - (d) Band structure of the phononic crystal whose rectangular waveguide has a width of $d = d_0$, $d_1$, $d_2$ with circulating airflow ($v = 10 m/s$, red curves) and without airflow (blue curves). The bands under the bandgap have different Chern numbers.

FIG. 3 (a) Band structure of a $1 \times 8$ supercell ($d = d_2$, $v = 10 m/s$) shows edge bands (colored lines) and bulk bands (grey lines). Red and blue lines represent different edge modes. (b) The pressure field distribution of edge modes for A: $k = -\frac{0.1\pi}{a}$ and B: $k = \frac{0.1\pi}{a}$.

FIG. 4. Demonstration of novel properties resulting from nontrivial bandgaps by tuning the geometric size of the phononic crystal. The source, indicated by a star, is a point source at $139 Hz$ frequency. (a) Topologically protected one-way propagation that is immune to defects and without backscattering. Here, the width of the waveguide is $d = d_2$ and the velocity of the airflow is $v = 10 m/s$. (b) One-way interface state propagation on the interface between two different lattices. The width of the waveguide is $d = d_1$ in the upper half and $d = d_2$ in the lower half. The arrows in (a) and (b) indicate the direction of propagation of the edge or interface mode.

FIG. 5 (a) The eigenfrequency of the eigenstates varies as a function of $v$. Black and red curves correspond to the $\varphi_s$ and $\varphi_{py}$ modes, respectively. The purple and blue areas indicate the region of the bandgap with a different topological invariant, where the intersection is the topological transition point. (b) The simulated pressure field distribution excited by a point source in a phononic crystal with airflow ($v = 5 m/s$). (c) The same as (b), but the velocity field of the airflow is $v = 15 m/s$. In both cases, the rectangular waveguide in the phononic crystal has a width of $d = 0.065 m$. The source has a frequency of $f = 138.4 Hz$ and is marked by a star. The arrow indicates the direction of propagation of the edge mode.

Table 1: Eigenvalues of the symmetry operators of eigenstates at different high symmetry points and resulting contributions to the gap Chern numbers. Subscripts 1 and 2 correspond to the bands with eigenfrequencies from low to high.

Figure 1

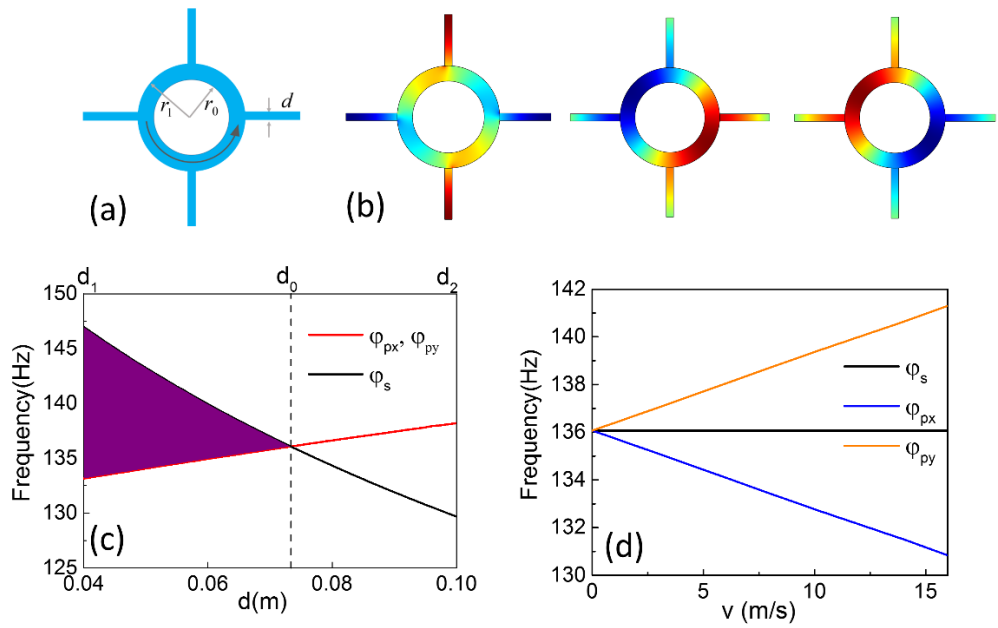

Figure 2

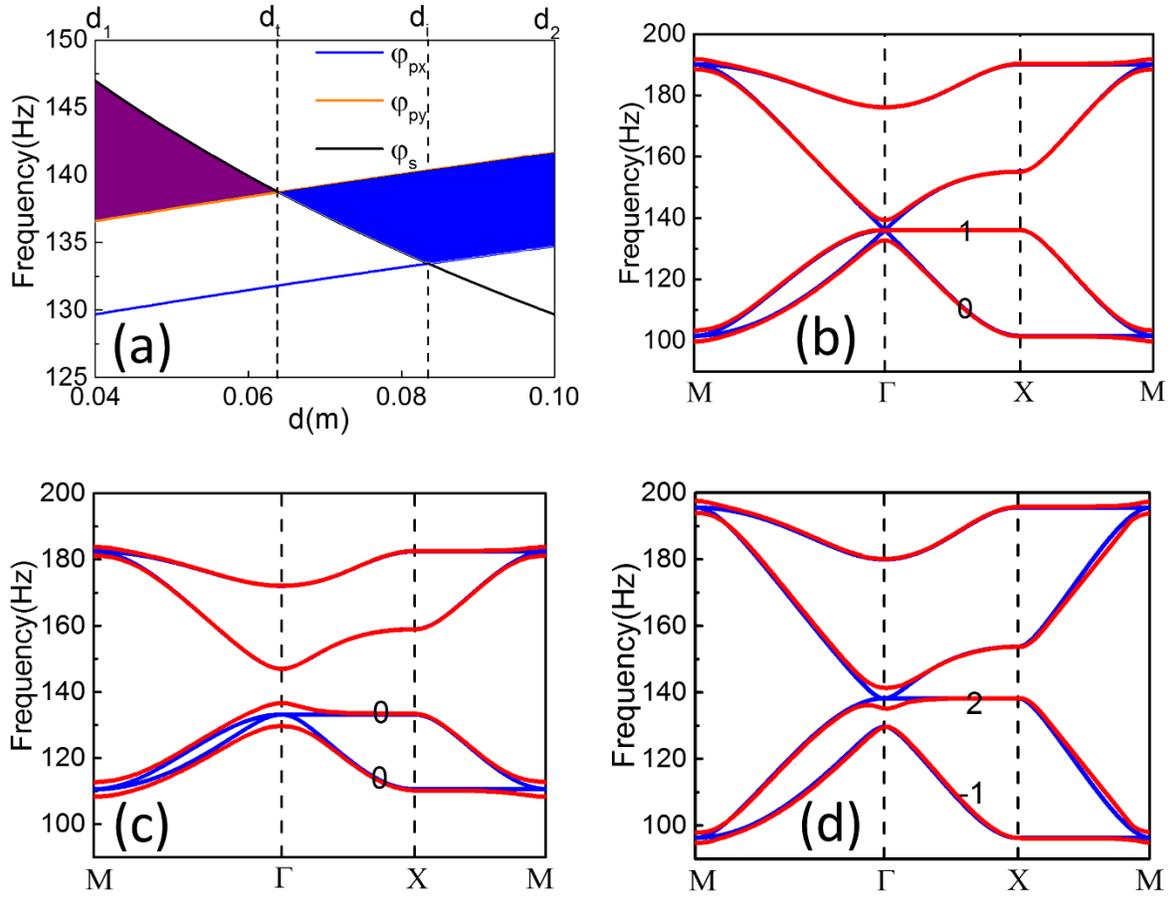

Figure 3

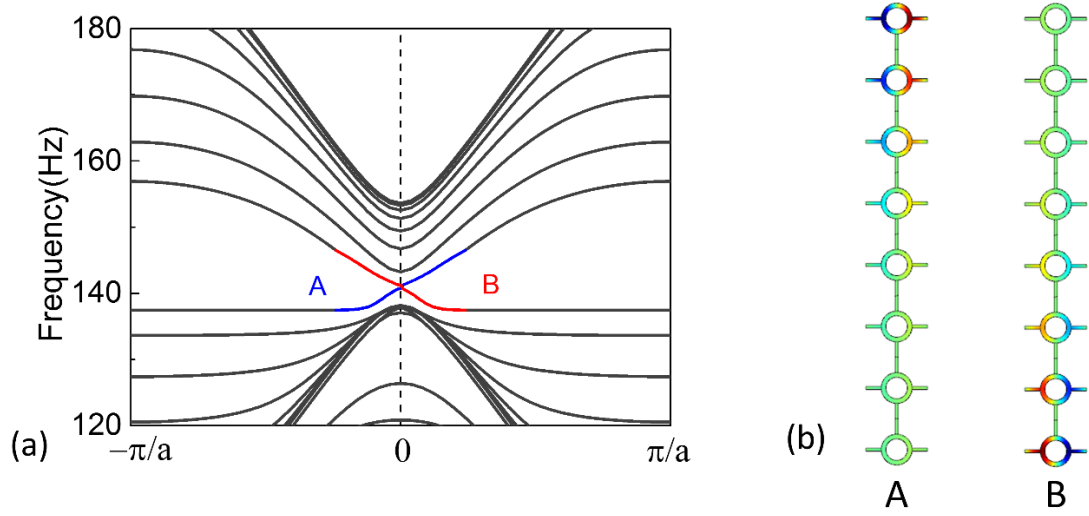

Figure 4

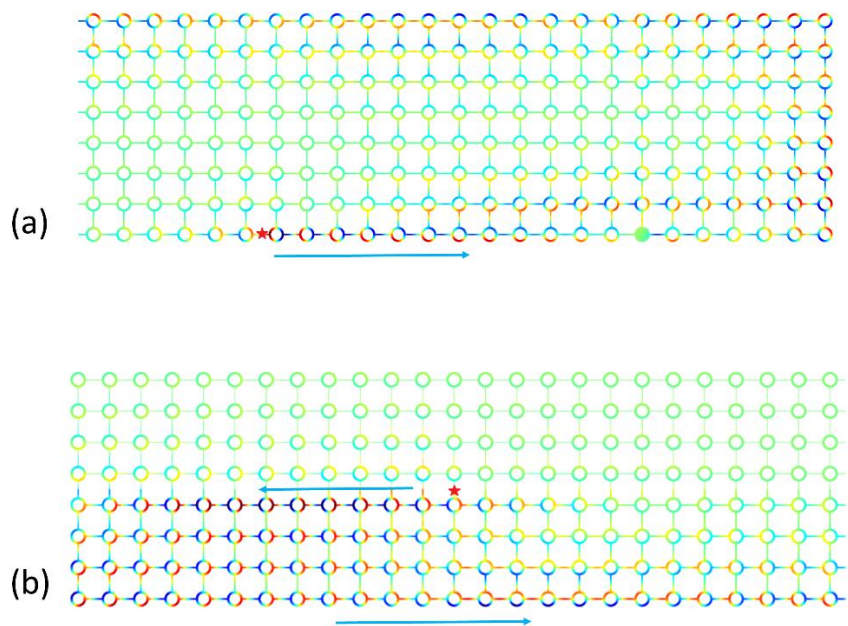

Figure 5

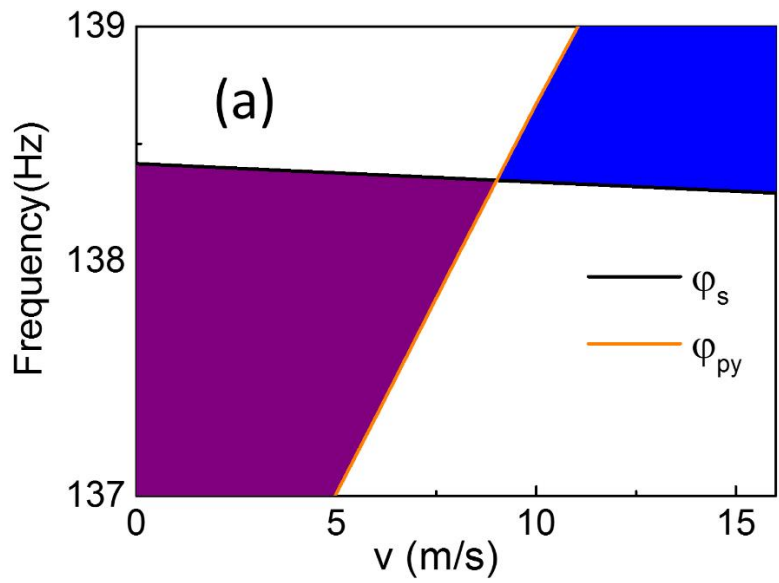

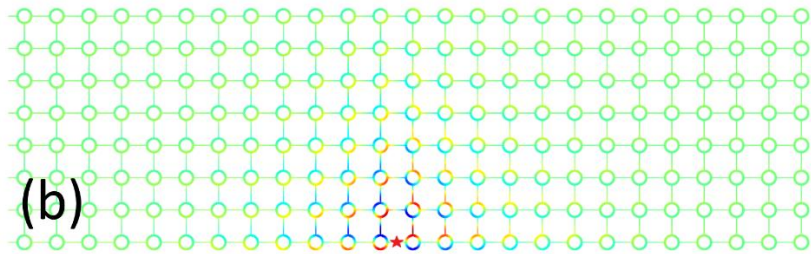

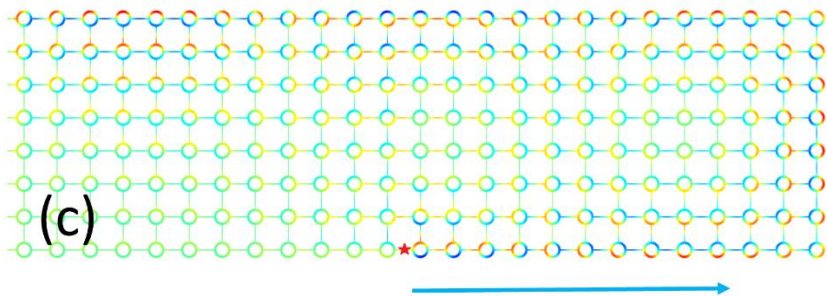

Table 1

|  | $d_1$ | $d_0$ | $d_2$ |
|---|---|---|---|
| $\hat{C}_4 \Gamma_1$ | $-i$ | $-i$ | $-1$ |
| $\hat{C}_4 M_1$ | $-i$ | $-i$ | $-i$ |
| $\hat{C}_2 Y_1$ | $-1$ | $-1$ | $-1$ |
| $\hat{C}_4 \Gamma_2$ | $i$ | $-1$ | $-i$ |
| $\hat{C}_4 M_2$ | $i$ | $i$ | $i$ |
| $\hat{C}_2 Y_2$ | $-1$ | $-1$ | $-1$ |
| $C_{gap}$ | 0 | 1 | 1 |